\documentclass[aps,prl,preprint,grouppedaddress,showpacs,showkeys]{revtex4}

\usepackage{graphicx}
\usepackage{dcolumn}
\usepackage{amsmath}
\usepackage{multirow}
\usepackage{epsfig}
\usepackage{amssymb}
\usepackage{color}

\begin{document}

\title{Revised symmetry rule and intrinsically time-reversal symmetry breaking pairing
in multi-orbital superconductors}

\author{Chang-Youn Moon}
\email{cymoon@kriss.re.kr}
\affiliation{Material Property Metrology Group, Korea Research Institute of Standards and Science, 
Yuseong, Daejeon 305-340, Republic of Korea}

\date{\today}

\begin{abstract}
We investigate the basic symmetry rule for the particle permutation in superconducting (SC)
pairing states by examining the numerical solution of the
linearized Eliashberg equation for Sr$_2$RuO$_4$. We find that
the general multi-band, frequency-dependent SC gap function does not simply transform to 
itself up to the minus sign with
either orbital ($\hat{O}$) or frequency ($\hat{T}$) exchange between two pairing electrons,
contradicting the common assumption which has been used without verification.
It originates from the fact that paring interactions are not invariant under the
$\hat{O}$ or $\hat{T}$ operation, and is demonstrated to be essential to correctly interpret and understand 
multi-band SC states. One of unique properties implied by our newly found symmetry rule
is the possibility of the complex eigenvalue in the linearized gap equation
where the corresponding gap functions $\Delta$ always have real and imaginary components both non-zero in
real frequencies, inherently breaking the time-reversal symmetry.
Our numerical results suggest the possibility of this unique pairing in a real material, Sr$_2$RuO$_4$,
for which the Hundness of the material is found to play a key role.
The revised symmetry rule not only leads us to more comprehensive understanding of the known SC
states, but also opens new possibilities into exotic and unique states.
\end{abstract}


\maketitle

The superconducting (SC) order parameter, $\Delta$, contains information on both the pairing wave
function and the excitation spectrum and hence is a key quantity to characterize
a SC state. It is often conveniently classified by its symmetry in the momentum space,
usually in the angular momentum eigenstates $s$, $p$, $d$, and so on, hinting at the underlying 
pairing mechanism. $\Delta$ has been further classified by whether it is even or odd under the exchange
of other possible degrees of freedom between the two pairing electrons, such as the spin (S), the 
time or the frequency (T), and the orbital (O), except for the parity (P) which is already included 
in the symmetry of $\Delta$ in the momentum space. When all the degrees of freedom are simultaneously
exchanged, $\Delta$ reverses its sign as dictated by the fermion exchange rule, which is compactly
expressed as $ \hat{S}\hat{P}\hat{O}\hat{T} = -1$ \cite{SPOT}, where the hat symbol represents the exchange
operator for the corresponding degree of freedom. Furthermore, it
has been assumed that $\Delta$ is an eigenfunction for each of the four exchange operators with the
eigenvalue $\pm1$, so that there are eight possible combinations of the eigenvalues which has been
considered as a standard symmetry classiﬁcation of the SC order parameter. Unconventional
superconductivity researches so far basically have relied on this symmetry classiﬁcation, and
it is especially relevant to some of exotic form of the superconductivity, such as the odd-
frequency pairing \cite{SPOT}. Meanwhile, the time-reversal symmetry (TRS)
is another important symmetry in superconductors, which is, when broken (i.e., $\Delta^*(-k)\neq\Delta(k)$ for
the spin singlet), associated with highly exotic 
SC phases having magnetizations and the non-trivial topology \cite{topoSC}.

As is usual for general unconventional superconductors, the pairing symmetry still remains elusive
for Sr$_2$RuO$_4$. The possibility of the $p$-wave spin triplet superconductivity
from the early stage of the research \cite{MackenzieMaeno,RiceSigrist,Sigrist,MaenoKittaka,Kallin}
has been found inconsistent with recent spin susceptibility measurements \cite{Pustogow,IshidaManago,Petsch}.
Instead, current discussions are mostly focused on more general two-component \cite{Benhabib,Ghosh}, 
TRS breaking \cite{Grinenko,Luke,Xia,Kapitulnik} chiral spin singlet order parameters. 
They are in the form of a complex combination among even parity order parameters such as $s$-, $d$-, and 
$g$-waves having either the symmetry-protected or the accidental degeneracy, with $s+id$ as an example. 
Theoretical/computational studies have often adopted
the linearized Eliashberg formalism, where the local electron-hole interaction, a crucial quantity in 
constructing the effective pairing interaction, is either approximated to be static in the random 
phase approximation (RPA) \cite{RomerScherer,RomerKreisel,RomerHirschfeld,RomerMaier,Gingras1,Gingras2} 
or frequency-dependent \cite{Kaser,moonSC}. The gap solutions are found to be sensitive to the
detail of a given calculation scheme, so that the proposed gap symmetries are diverse among the studies.
Therefore, instead considering just the magnitude of the eigenvalue of a gap solution,
it is desirable to take into account more comprehensive physical properties of the gap solution 
to decide its significance \cite{Kaser}.

In this work, we solve the frequency-dependent linearized Eliashberg equation within the density-functional 
theory combined with the dynamical mean-field theory (DFT+DMFT) framework. Our numerical results show 
that the symmetry property of a pairing state $\Delta$ with respect to
the $\hat{O}$ or $\hat{T}$ operation does not necessarily follow previously used symmetry rule, 
$\hat{O} \Delta=\pm\Delta$ and $\hat{T}\Delta=\pm\Delta$, which has been assumed to hold without 
a strict justification. We revisit 
the basic symmetry rule on the permutation of a degree of freedom between two pairing fermions, 
and conclude
that the general pairing interaction is not invariant under $\hat{O}$ or $\hat{T}$ operation but transforms
to its complex conjugate in the imaginary frequency, leading to our results on the symmetry of $\Delta$ 
as observed. The importance of our revised symmetry rule is demonstrated on correctly interpreting 
the computational results from linearized gap equations. Meanwhile, we notice an unique type
of the gap solutions associated with a complex eigenvalue in the gap equation, which is found to be intrinsically 
TRS breaking when analyzed using our suggested symmetry rule. We show that this exotic pairing
state can be realized in materials with strong Hund's coupling, such as Sr$_2$RuO$_4$. Our
work re-establishes the basic symmetry properties of the SC order parameter, 
providing an insight into new forms of pairing states as well as more complete understanding 
of conventional ones.

We use the modern implementation of DFT+DMFT method within all electron embedded
DMFT approach \cite{DMFT}. This is a parameter-free ﬁrst-principles method in the non-interacting
electron limit, while local electron correlation eﬀects are described with two parameters, the
intra-orbital Coulomb repulsion $U$ and the Hund's coupling $J$.
Internal atomic positions are optimized, and $U = 4.5$ eV and $J = 1.0$ eV 
are adopted in the Slater parametrization consistently with previous
studies on this material \cite{DengHaule}. All calculations are done at 174 K. 
Spin-orbit coupling (SOC) can have
non-negligible effects on the electronic structure mainly near the region where
different sheets of FS intersect, by strong orbital mixing \cite{ZhangGorelov,Kim,Tamai}. 
As evaluating the two-particle vertex including SOC is still not available, and also 
considering susceptibilities are less affected by SOC than one-particle spectra \cite{Gingras1},
we neglect SOC in this work. 
%Other details of calculations are depicted in the Supplemental 
%Material \cite{SM}. 

We start with a brief review on the linearized Eliashberg formalism in DFT+DMFT \cite{moonSC,Yin2014}.
To set up the gap equation, first we need to compute spin (m) and charge (d) susceptibilities
by combining the polarization bubble $\chi_0$ and the two-particle vertex $\Gamma^{m/d}$
in the Bethe-Salpeter equation:
\begin{equation}
\chi^{m/d}_{\alpha\alpha';\beta\beta'}(i\nu,i\nu')_{\Omega, q} = ((\chi^{0}_{\Omega,q})^{-1} - \Gamma^{m/d}_\Omega)^{-1}_
{\alpha\alpha';\beta\beta'}(i\nu,i\nu'),
\end{equation}
where $\alpha^{(')}$ and $\beta^{(')}$ are orbital
indices, $\nu^{(')}$ and $\Omega$ are fermionic and bosonic Matsubara frequencies, respectively.
Here $\chi_0$ is obtained by multiplying two
one-particle Green's functions which contain the local correlation effect within the DFT+DMFT
framework, while $\Gamma^{m/d}$ is evaluated in the CTQMC impurity solver and includes
the local correlation effect in the two-particle level. $\Gamma^{m/d}$ is fully dynamic in
our calculation which has been found to be crucial to reproduce the correct peak position
of the spin susceptibility in the $q$-space as observed in the experiments \cite{moonSC}.
Then we can construct the linearized Eliashberg equation:
\begin{equation}
 -k_BT\sum_{k'\nu'\alpha'\beta'\gamma\delta}\Gamma^{pp,s/t}_{\alpha\beta;\alpha'\beta'}(k,i\nu;k',i\nu')
\chi^{0,pp}_{\alpha'\beta';\gamma\delta}(k,'i\nu')\Delta_{\gamma\delta}(k',i\nu')=\lambda\Delta_{\alpha
\beta}(k,i\nu),
\end{equation}
where $\Gamma^{pp,s/t}$ is the two-particle vertex in the particle-particle and spin singlet (s)/triplet (t)
channels serving as the effective pairing interaction and consists of $\Gamma^{m/d}\chi^{m/d}\Gamma^{m/d}$ 
in appropriate combinations of the indices of degrees of freedom. In this study,
we only consider the spin singlet case in accordance with experimental observations. 
The eigenfunction $\Delta$ is interpreted as the gap function while the dimensionless eigenvalue $\lambda$
represents the pairing strength with the maximum possible value of 1. This equation is derived
as the condition for the static superconducting susceptibility defined in the normal state to diverge 
as the temperature decreases towards the transition temperature $T_c$, 
at which $\lambda$ reaches 1. Above $T_c$, $\lambda$ is smaller than 1 and the susceptibility 
is finite, representing the system is in the normal state. Therefore, the solutions $\Delta$ of 
the equation with $\lambda$ smaller than 1 at a given temperature, as usually obtained in our study 
as well as previous works adopting similar computational methods, do not correspond to true 
superconducting states with the static long-range order as expected below $T_c$. Instead, 
they represent superconducting fluctuations whose strength is reflected in the
magnitude of $\lambda$ in the normal state. Nevertheless, a strong fluctuation
with relatively larger $\lambda$ above $T_c$ is more likely to eventually 
lead to a static superconducting order as the temperature drops. Other 
details of the formalism are described in previous studies \cite{moonSC,Yin2014}. 

The eigenfunction $\Delta$ in eq. (2) is approximated to be constant with respect to 
the imaginary frequency $i\nu$ in our previous study \cite{moonSC}, constraining
$\Delta$ to be even with respect to the $\hat{T}$ operation, i.e., $\hat{T}\Delta(i\nu)=\Delta(-i\nu)=
\Delta(i\nu)$. On the contrary, our current study has no such constraint, 
and $\Delta$ can be even, odd, or something else as we will see later.
In Fig. 1, we categorize our numerical solutions calculated for Sr$_2$RuO$_4$ into three distinct
types in terms of the symmetry with respect to the $\hat{T}$ operation.
The first category is for the intra-orbital pairing, as shown with the gap function $\Delta(k)$ 
in the momentum space in the orbital basis in Fig. 1(a). We can notice $\Delta$ is an odd function in the k-space,
$\hat{P}\Delta(k)=\Delta(-k)=-\Delta(k)$. 
Also, the intra-orbital solution is an even function by definition for the $\hat{O}$ operation, 
$\hat{O}\Delta_{\alpha\alpha}=\Delta_{\alpha\alpha}$. Then, the only possibility left for this solution for the $\hat{T}$ operation
is that it is an odd function, $\hat{T}\Delta(i\nu)=
\Delta(-i\nu)=-\Delta(i\nu)$, to satisfy $\hat{S}\hat{P}\hat{O}\hat{T}=-1$. It is indeed
found to be the case as shown in Fig. 1(a). Therefore, in this symmetry category
for the intra-orbital pairing, $\Delta$ is either even or odd with $\hat{T}$.
Meanwhile, our numerical results show that the symmetry property is not so straightforward 
for inter-orbital gap solutions $\Delta_{\alpha\beta}$ where $\alpha \neq \beta$,
with an example displayed in Fig. 1(b) as the second category. In this specific example, $\hat{O}\Delta_{\alpha\beta}
=\Delta_{\beta\alpha}=\Delta_{\alpha\beta}^*$ and $\hat{T}\Delta(i\nu)=\Delta(-i\nu)=
\Delta^*(i\nu)$. The gap function is not even nor odd under each of $\hat{O}$ and $\hat{T}$ operations,
but transforms to its complex conjugate in the imaginary frequency domain.
However, the combined operation $\hat{O}\hat{T}=1$ also in this case, satisfying the universal
relation $\hat{S}\hat{P}\hat{O}\hat{T}=-1$ for the fermion permutation. 
Finally, we find inter-orbital solutions can have even more unexpected $\hat{O}/\hat{T}$ symmetry,
as introduced in Fig. 1(c) and (d) as the third category.
They are doubly degenerate solutions with a complex eigenvalue $\lambda$, accompanied by another 
degenerate pair of solutions with the eigenvalue $\lambda^*$. These gap functions have 
essentially decoupled positive and negative imaginary frequency components, with the weight
mainly on either sign of the frequency component. Instead of having a definite relation between
the positive and negative components in a given gap solution, they are complex conjugate
to each other between a solution with $\lambda$ and another solution with $\lambda^*$.
Our results in Fig.1(b)-(d) indicate that inter-orbital gap functions are
eigenfunctions of neither $\hat{O}$ nor $\hat{T}$ operation, which is in contradiction to
the common assumption that a gap function is either even or odd under each of the $\hat{S}\hat{P}\hat{O}\hat{T}$
operations. To resolve this seemingly contradictory results, we revisit the basic symmetry properties
of the SC gap functions under the $\hat{S}\hat{P}\hat{O}\hat{T}$ operations.

When the gap equation eq. (2) is expressed in a simpler form as an eigenvalue equation, $V\Delta=\lambda\Delta$
with $V$ as the effective pairing interaction,
it is obvious that $\hat{S}^\dag V\hat{S}=V$ when $V$ is in the spin singlet or the triplet basis 
resulting in $\hat{S} \Delta_{s/t}= \pm \Delta_{s/t}$ where plus/minus sign is for the spin triplet/singlet state.
Also, $\hat{P}^\dag V\hat{P}=V$ for systems with the inversion symmetry, so that $\hat{P} \Delta=\pm\Delta$.
Meanwhile, it is not very obvious whether $V$ is invariant under the orbital or frequency exchange. For
the intra-orbital pairing for which $\hat{O}$ is the identity operator, the fundamental relation $\hat{S}\hat{P}\hat{O}\hat{T}=-1$
guarantees that $V$ is also invariant under the $\hat{T}$ operation, so that $\Delta$ is either
an even or odd function of the frequency with an example demonstrated in Fig. 1(a). 
However, for the inter-orbital pairing, 
there is no general reason why $V$ should be invariant under each of the orbital and frequency exchanges,
although it is invariant for the combined operation $\hat{O}\hat{T}$. 
Our key finding is that the matrix $V$ is not invariant under the operations 
but transforms to its complex conjugate in the imaginary frequency domain 
\begin{equation} 
	\hat{O}^\dag V \hat{O}=\hat{T}^\dag V \hat{T}=V^*.
\end{equation}
Then we can show that 
\begin{equation}
	\sum_{\gamma\delta}V_{\alpha\beta;\gamma\delta}\Delta^*_{\delta\gamma}=\lambda^*\Delta^*_{\beta\alpha} \qquad    \mathrm{and} 
	\qquad    \sum_{\nu'}^{}V(i\nu,i\nu')\Delta^*(-i\nu')=\lambda^*\Delta^*(-i\nu),
\end{equation}
where the matrix multiplication between $V$ and $\Delta$ over the other unspecified degrees of freedom is also implied. 

The consequences of these relations
can be classified by the complexity of the eigenvalue $\lambda$ of the gap equation, eq. (2). 
When $\lambda=\lambda^*$, i.e., $\lambda$
is real, as is for most of cases, eq. (4) leads to 
\begin{equation}
	\hat{O}\Delta_{\alpha\beta}=\Delta_{\beta\alpha}=e^{i\theta_1}\Delta^*_{\alpha\beta}
	\qquad {\rm and} \qquad \hat{T}\Delta(i\nu)=\Delta(-i\nu)=e^{i\theta_2}\Delta^*(i\nu).
\end{equation}
where the arbitrary constant phase angles, $\theta_{1/2}$, are
related to be $e^{i(\theta_1-\theta_2)}=\pm1$ for $\hat{S}\hat{P}\Delta=\mp\Delta$ to satisfy $\hat{S}\hat{P}\hat{O}\hat{T}\Delta=-\Delta$.
When we choose the phase factor $e^{i\theta_2}$ to be 1 without 
loss of generality, we have
%the $\hat{T}$ operation transforms the gap function in the imaginary frequency 
%to its complex conjugate. Then the other phase factor $e^{i\theta_1}$ is determined accordingly to be $\pm1$
%depending on the sign of the $\hat{S}\hat{P}$ operation on the gap function as mentioned above, so that
%the $\hat{O}$ operation also transforms the gap function to its complex conjugate up to the minus sign:
$\Delta_{\beta\alpha}=\pm\Delta_{\alpha\beta}^*$ and $\Delta(-i\nu)=\Delta^*(i\nu)$
for $\hat{S}\hat{P}\Delta=\mp\Delta$.
The gap function shown in Fig. 1(b) belongs to this category. 
%Again in the intra-orbital case, $\hat{O}^\dag V \hat{O}=\hat{T}^\dag V \hat{T}=V$ should also hold as mentioned 
%earlier, then $\Delta_{\alpha\alpha}(-i\nu)=\Delta_{\alpha\alpha}(i\nu)$ or $-\Delta_{\alpha\alpha}(i\nu)$,
%and at the same time, $\Delta_{\alpha\alpha}(-i\nu)=\Delta_{\alpha\alpha}^*(i\nu)$ or $-\Delta_{\alpha\alpha}*(i\nu)$ from
%eq. (4). This implies that $\Delta_{\alpha\alpha}$ is either purely real or imaginary. 
Here one should note that eqs. (3) - (5) are valid only in the imaginary frequency domain. Meanwhile, there is no such general relation 
of $V$ at all in the real frequency domain, and it is simply
$\hat{O}^\dag V \hat{O}$ (or $\hat{T}^\dag V \hat{T}) \neq V$, hence $\hat{O}$(or $\hat{T})\Delta$ is not $\pm\Delta$
nor $\pm\Delta^*$. Therefore, it is concluded that for the general multi-orbital superconductivity, the order parameter $\Delta$
has no relation between $\Delta_{\alpha\beta}(\omega)$ and $\Delta_{\beta\alpha}(\omega)$, or between 
$\Delta_{\alpha\beta}(\omega)$ and $\Delta_{\alpha\beta}(-\omega)$ for real frequency $\omega$.
This is also easy to see by considering a schematic as displayed in Fig. 2(a). For two general non-degenerate
bands (orbitals) denoted by $\alpha$ and $\beta$, the propagation of an electron pair at ($\alpha,k,\omega$) and
($\beta,-k,-\omega$) is not equivalent to that at ($\alpha,k,-\omega$) and ($\beta,-k,\omega$). 

Previous theoretical studies have classified their calculated gap solutions into even/odd functions 
of orbital ($\hat{O}$) and frequency ($\hat{T}$) \cite{Gingras1,Gingras2,Kaser} which is now turned out to be inappropriate for
both imaginary and real frequency domains. It is crucial to strictly keep the complex-valuedness 
of the inter-orbital elements in the pairing potential $V$ in the gap equation.
$V$ can be mistaken to be invariant under $\hat{O}$ or $\hat{T}$, $\hat{O}^\dagger V \hat{O}=\hat{T}^\dagger V \hat{T}=V$, 
when only the real components of $V$ are retained in the calculation, or an artificial
symmetry is imposed on $V$, for examples, leading to the artificial even-odd behavior of the gap function. 
Moreover, the lowest Matsubara (imaginary) frequency component of the gap function
has been often considered to
be approximately identical to its zero real frequency component, $\Delta(\omega=0) \sim \Delta(i\nu_0)$
which contains both real and complex components in general.
However, eq. (5) suggests that a care must be taken to estimate $\Delta(\omega=0)$ depending on the
overall phase factors of the gap functions. For example, for $\Delta(-i\nu)=\Delta^*(i\nu)$ with the 
choice of the phase factor $e^{i\theta_2}=1$, the imaginary component should go to zero towards
the zero frequency, indicating that only the real part of $\Delta(i\nu_0)$ should be taken, 
$\Delta(\omega=0) \sim {\rm Re}[\Delta(i\nu_0)]$.

Usually, the real frequency function can be obtained from its imaginary frequency counterpart by using the analytic
continuation, which is not available in this study. However, we can derive
some of useful information of the real frequency $\Delta$
by simply considering the Fourier transformation. For the zero (equal) time pairing for which real and imaginary times are identical,
\begin{equation}
	\sum_{\nu}\Delta(i\nu)=\Delta(\tau=t=0)=\int_{-\infty}^{\infty}\Delta(\omega)d\omega.
\end{equation}
Then, for a choice of the phase factor $e^{i\theta_2}=1$ leading to $\Delta(-i\nu)=\Delta^*(i\nu)$, 
we can see that Re$[\Delta(t=0)]\neq 0$ and Im$[\Delta(t=0)] = 0$.
Because there is no definite relation for an inter-orbital gap function between its positive and negative real frequencies
as explained earlier, neither even nor odd,
we can conclude that Re$[\Delta(\omega)]\neq 0$ and Im$[\Delta(\omega)]=0$ for general $\omega$. 
Therefore, the general inter-orbital gap function $\Delta(\omega)$
is purely real or can be transformed to be real by a constant overall phase factor, and TRS is preserved
as expected.

A rare but more interesting case, which our gap solutions introduced in Fig. 1(c)/(d) correspond to,
is when $\lambda$ is not equal to $\lambda^*$, i.e., $\lambda$ is a complex
number. %First we need to make sure that $\lambda$ can be a complex number within the linearized Eliashberg formalism.
%The matrix $V$ in the simplified form of an eigenvalue equation, $V\Delta=\lambda\Delta$, corresponds to 
%$\Gamma^{pp}\chi^{pp}$ as shown in eq. (2). While $\Gamma^{pp}$ represents the effective scattering amplitude 
%between the electron pair and is Hermitian, $\chi^{pp}$ is not, and hence $V$ is not Hermitian so that
%it is possible that $\lambda$ is complex in principle. 
When $\Delta_1$ is the eigenfunction for the eigenvalue
$\lambda$, then eq. (4) guarantees the existence of another solution $\Delta_2$ with the eigenvalue $\lambda^*$:
\begin{equation}
	\begin{aligned}
	V\Delta_1=\lambda\Delta_1 \qquad &\mathrm{and} \qquad V\Delta_2=\lambda^*\Delta_2 \\ \\
	\Delta_{1\beta\alpha}=e^{i\theta_1}\Delta_{2\alpha\beta}^* \qquad &\mathrm{and} \qquad \Delta_1(-i\nu)=e^{i\theta_2} \Delta_2^*(i\nu) 
%	\Delta_1=
%	\begin{pmatrix}
%		\vdots \\
%		\begin{aligned}
%		&\Delta_1(i\nu_2)  \\
%		&\Delta_1(i\nu_1)  \\
%		&\Delta_1(-i\nu_1) \\
%		&\Delta_1(-i\nu_2) \\
%	        \end{aligned}
%		\\
%		\vdots
%	\end{pmatrix}
%	\qquad &\mathrm{and}  \qquad
%	\Delta_2=\pm
%	\begin{pmatrix}
%		\vdots \\
%		\begin{aligned}
%		&\Delta_1^*(-i\nu_2)  \\
%		&\Delta_1^*(-i\nu_1)  \\
%		&\Delta_1^*(i\nu_1)   \\
%		&\Delta_1^*(i\nu_2)   \\
%	        \end{aligned}
%	        \\
%		\vdots
%	\end{pmatrix}
        \end{aligned}
\end{equation}
where the arbitrary constant phase angles, $\theta_{1/2}$, are
also related to be $e^{i(\theta_1-\theta_2)}=\pm1$ for $\hat{S}\hat{P}\Delta=\mp\Delta$ as in the real $\lambda$
case in eq. (5).
The opposite frequency sign components have no definite relation between each other 
within a given gap function $\Delta_1$ or $\Delta_2$ in contrast to the real $\lambda$ case, but only between $\Delta_1$ and $\Delta_2$.
Then, from eq. (6), we have Re$[\Delta_{1/2}(t=0)]\neq 0$ and 
Im$[\Delta_{1/2}(t=0)] \neq 0$, leading to Re$[\Delta_{1/2}(\omega)]\neq 0$ and Im$[\Delta_{1/2}(\omega)]\neq0$. 
%Additionally,
%it is obvious that $\Delta_1(t=0)=\pm\Delta_2^*(t=0)$, naturally pointing to the relation $\Delta_1(\omega)=\pm\Delta_2^*(\omega)$
%considering that no definite relation is given among the different real frequency components. 
These gap solutions have real and imaginary components both non-zero and independent each other on the real frequency axis. 
The gap functions then cannot be transformed to be purely real
by multiplying any phase factor $e^{i\theta}$, and hence are intrinsically TRS breaking, even without forming a complex
linear combination of degenerate gap functions in the form of $\Delta_a+i\Delta_b$.
As is true for usual real valued $\lambda$ solutions, these paired complex eigenvalues $\lambda/\lambda^*$ are to
approach the real value 1 as the decreasing temperature for their eigenfuctnions $\Delta_{1/2}$ to be actually realized 
as superconducting states. Then $\Delta_{1/2}$ become degenerate solutions with the same real eigenvalue 
$\lambda=\lambda^*=1$, and even though $\lambda$ is real the degenerate solutions still satisfy eq. (7) 
leading to the TRS breaking in the real frequency.

Our numerical solutions with a complex $\lambda$, as shown in Fig. 1(c)/(d), exhibit an unique frequency 
dependence where $\Delta_{1/2}$
have either positive or negative imaginary frequency components only, indicating the positive and negative frequency components
are decoupled in the effective pairing potential matrix $V$ which can naturally result in the relations
in eq. (7) \cite{SM}. RPA level analysis suggests that $J/U$, Hund's coupling strength compared with
the Coulomb repulsion, is a key to determine the opposite-frequency
decoupling strength in $V$ which is found to be essential for the complex $\lambda$ in our specific solutions \cite{SM}.
To demonstrate the relation
between $J/U$ and the frequency decoupling, we solve eq.(2) using different values of $J$ and $U$.
Figure 2(b) and (c) show the complex eigenvalue of the gap solution in Fig. 1(c)/(d) as a function of $J$ and $U$.
%Here only two-particle vertex functions $\Gamma^{m/d}$ are recalculated with varying $J$ and $U$ while 
%one-particle Green's functions are kept fixed with the default $J$ and $U$ values, 1.0 and 4.5 eV.
According to the RPA level analysis as discussed above, increasing $J$ drives $\Gamma^m_{\alpha\beta;\alpha\beta}=U'=U-2J$ 
more negative enhancing the frequency decoupling associated with the complex eigenvalue \cite{SM}. 
Indeed, the imaginary part of the eigenvalue $\lambda$, Im[$\lambda$],
increases with increasing $J$, while it becomes zero for a sufficiently small $J$ accompanying
a sudden increase of the real part in Fig. 2(b), indicative of the onset of the coupling between the opposite
frequency components of the pairing potential. Meanwhile, Im[$\lambda$] also initially increases with 
increasing $U$ because the frequency-diagonal component of the pairing potential $V(i\nu,i\nu)$ itself is 
enhanced by $U$. However, the frequency decoupling is eventually suppressed for larger $U$ 
with more positive $\Gamma^m_{\alpha\beta;\alpha\beta}=U'=U-2J$ within RPA, and this effect starts to 
dominate around $U=4.9$ eV with decreasing Im[$\lambda$] for larger $U$ until $\lambda$ becomes
real at $U=5.1$ eV. There is also a jump of Re[$\lambda$] at this $U$ as shown in Fig. 2(c),
pointing to the onset of the opposite frequencies coupling similarly with the small $J$ case in
Fig. 2(b). Therefore, it is indeed verified that the complex eigenvalue $\lambda$ solutions exist 
only for large $J$ and/or small $U$ values.

Our complex eigenvalue pairing states in Fig. 1(c)/(d) have the $f$-wave $E_u$ irreducible representation in the
$k$-space consisting
of doubly degenerate solutions. Any linear combination between the two degenerate states 
would be consistent with the two-component nature of the order parameter in Sr$_2$RuO$_4$ suggested by ultrasound
measurements \cite{Benhabib,Ghosh}, while a single component alone could form a nematic SC
state with the orthorhombic symmetry instead of the original tetragonal symmetry \cite{Benhabib}.
Previously suggested pairings are TRS preserving if not in the form of the complex linear combination 
between two degenerate states, so that, for example, a single $E_g$ component ($d_{xz}$ or $d_{yz}$)
nematic pairing does not break TRS \cite{Maeno2024}. On the contrary, our states are always TRS breaking,
either in a single component or in a two-component form, because each of the degenerate states is 
TRS breaking with the complex gap function as discussed earlier.
One possible way to unequivocally identify the realization of the complex eigenvalue pairing then would be 
the observation of a single-component nematic SC state which is also TRS breaking, 
a combination of previously suggested SC features which is not accessible from conventional TRS preserving gap 
function components.
While the actual feasibility of the complex eigenvalue pairing in Sr$_2$RuO$_4$ is still inconclusive with 
the experimental evidences at hand, this unique form of the SC pairing is worth the attention
in general SC materials especially with the significant role of the Hund's coupling
in the electronic correlation such as Hund's metals.

In summary, we solve the frequency-dependent linearized Eliashberg equation within the DFT+DMFT 
method to investigate the nature of
the SC pairing in Sr$_2$RuO$_4$ and also of more general multi-orbital SC systems. 
We argue that the pairing interaction is not invariant under the frequency nor the orbital exchange, 
and the gap functions are neither even nor odd under these operations in contrast to the basic
assumption in the widely accepted ''SPOT'' symmetry classification of the SC order parameter.
As an unique case of the non even-odd behavior in the frequency/orbital exchange of the inter-orbital gap function,
we show that a pair of gap solutions can emerge with eigenvalues complex conjugate to each other. 
These gap functions have
real and imaginary components both non-zero in the real frequency domain, hence inherently breaking TRS. 
We also demonstrate the specific realization of this type of pairing state in Sr$_2$RuO$_4$ is associate with
the strong Hund's coupling of the material.
Our work provides an important insight to extend our knowledge on the basic symmetry relations of the
SC order parameter, from which the possibility of an unique and exotic pairing state is aslo found 
and understood.

\begin{acknowledgments}
\end{acknowledgments}

\newpage

\begin{figure}[tp]
\includegraphics[width=0.8\linewidth]{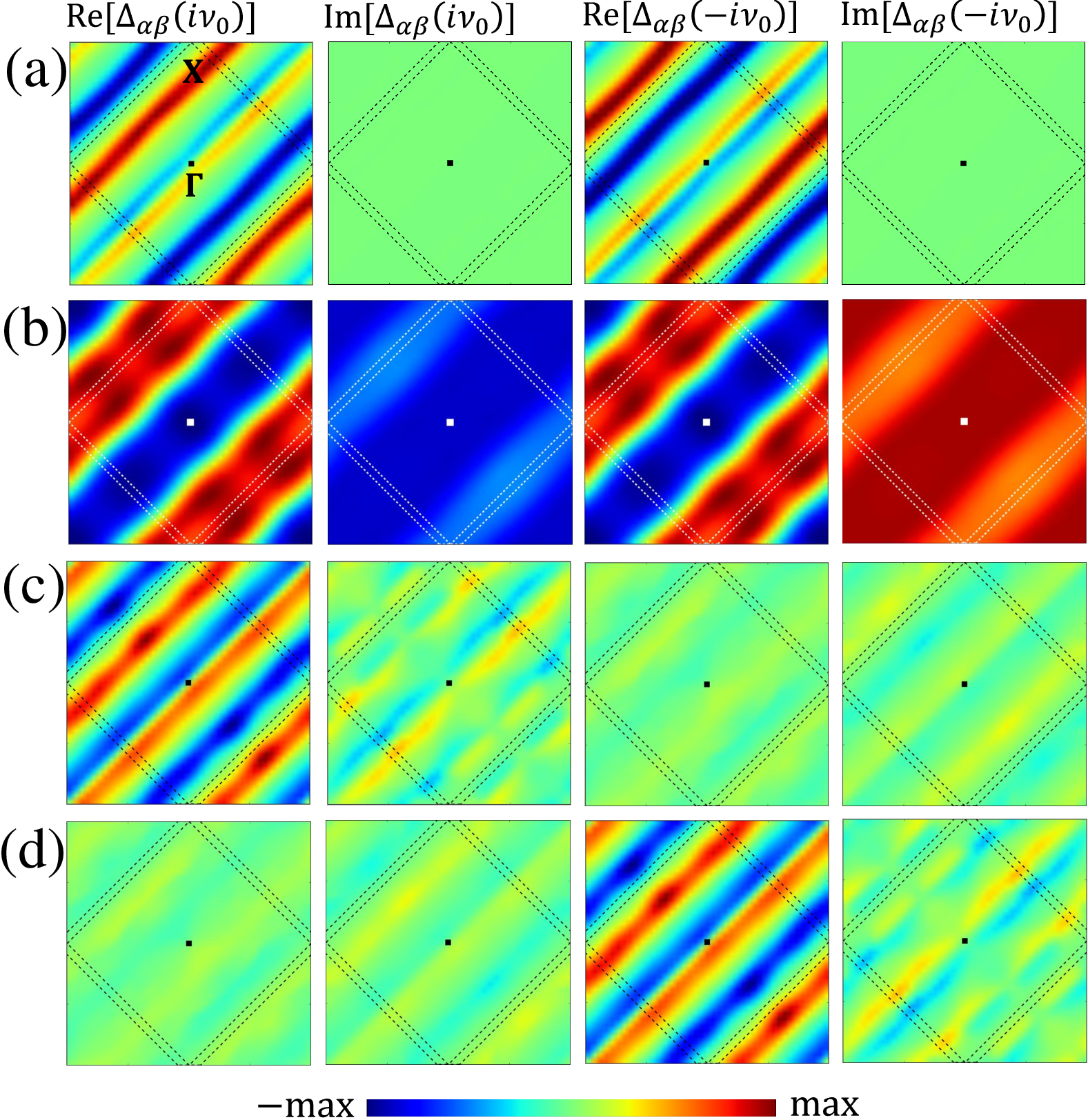}
\caption{Examples of our numerical solutions for the SC gap function $\Delta_{\alpha\beta}(k,i\nu)$ 
	on the $k_z=0$ plane, in the orbital basis $\alpha\beta$, at the first positive ($i\nu_0$) and negative 
	($-i\nu_0$) Matsubara frequencies, obtained for Sr$_2$RuO$_4$. Both real (Re) and imaginary (Im) parts are shown. 
	(a) an intra-orbital ($\alpha=\beta=d_{xz}$) solution
	where $\Delta(i\nu)=-\Delta(-i\nu)$, (b) a general inter-orbital ($\alpha=d_{xz}$, $\beta=d_{xy}$)
	solution with a real eigenvalue $\lambda$ where $\Delta(i\nu)=\Delta^*(-i\nu)$, and (c) a special 
	inter-orbital ($\alpha=d_{xz}$, $\beta=d_{xy}$) solution with a complex eigenvalue $\lambda$ where 
	there is no general relation between $\Delta(i\nu)$ and $\Delta(-i\nu)$, along with (d) the associated
	solution with $\lambda^*$. For (c) and (d), $\Delta_1(i\nu)=\Delta_2^*(-i\nu)$ where $\Delta_{1/2}$
	is the eigenfunction for the eigenvalue $\lambda/\lambda^*$. In case of (a), $\Delta(i\nu)=-\Delta(-i\nu)$ indicates
	that $\Delta(\omega=0)=0$, while $\Delta(\omega=0) \sim \mathrm{Re}[\Delta(i\nu_0)]$ for (b) since $\Delta(i\nu)=\Delta^*(-i\nu)$.
}
\label{fig1}
\end{figure}

\begin{figure}[tp]
\includegraphics[width=0.8\linewidth]{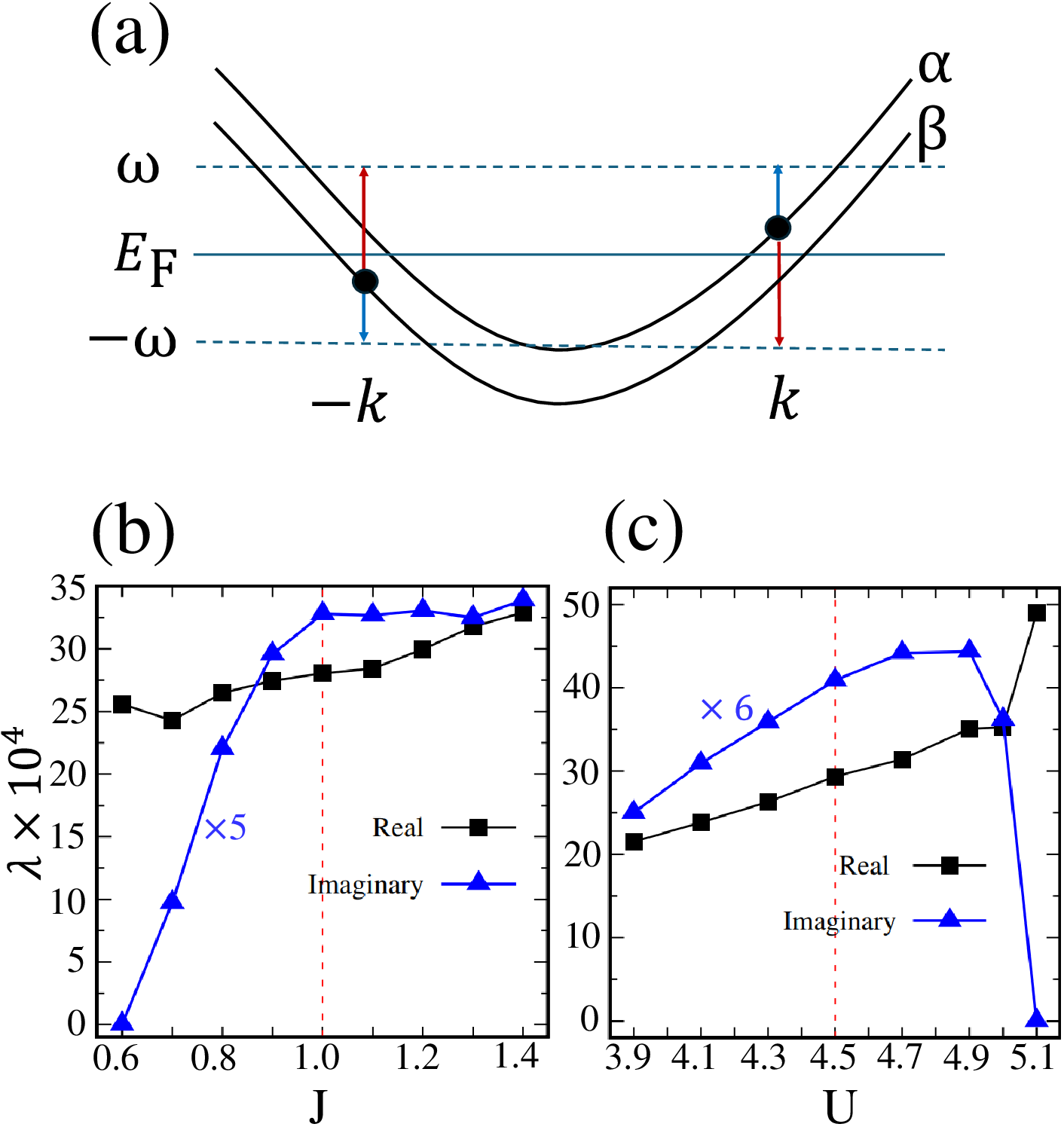}
\caption{(a) A schematic showing the inequivalence of the electron pair propagation between
	$(\alpha,k,\omega)$/$(\beta,-k,-\omega)$ and $(\alpha,k,-\omega)$/$(\beta,-k,\omega)$.
	It is clear that the inter-orbital pairing interaction is not invariant under either the frequency ($\hat{T}$)
	nor orbital ($\hat{O}$) exchange operation, while invariant for the simultaneous exchange 
	of the two degrees of freedom. For the intra-band pairing, $\alpha=\beta$,
	the pairing interaction is invariant under the $\hat{T}$ operation.
	The complex eigenvalue $\lambda$ of the gap solution as shown in Fig. 1(c), with the imaginary part five or six times 
	enhanced, is displayed as
	a function of (b) the Hund's coupling $J$ and (c) the intra-orbital Coulomb repulsion $U$. With
	decreasing $J$ from $J=1$ eV, the imaginary part of $\lambda$ monotonically decreases to zero at $J=0.6$,
	while it is saturated over $J=1$ eV. In case of $U$, Im[$\lambda$]
	initially increases with $U$ but eventually goes to zero at a higher $U$ value, 5.1 eV.
	In both cases, sudden upturns of the real part of $\lambda$ accompany the suppression of the 
        imaginary part to zero. The standard values of $U$ and $J$ adopted in this study with which the results
        in Fig. 1 are obtained, 4.5 eV and 1.0 eV, respectively, are marked by the vertical dashed lines.
	$U$ and $J$ are fixed to 4.5 eV and 1.0 eV for (b) and (c), respectively.
} 
\label{fig2}
\end{figure}

\end{document}